# SN1a Supernova Red Shifts

R.L. Collins retired U.T. Austin

## Abstract

Recent SN1a data have probed deeper into space than ever before. Plotted as distance vs. recession speed, a disturbing non-linearity is found which has led to speculations about "dark energy" which somehow acts like anti-gravity. This study finds a full explanation in relativity theory. The metric of space shrinks, in the presence of a gravitational potential, V, by $\exp(V/c^2)$. Early in the big bang, when the SN1a's sent their signals, V was larger than now. By fitting the data to a relativistic model, we probe the metric of the early big bang. V, due to all mass in the big bang, is a billion times larger than that provided by earth gravity alone. The big bang is modeled as a sphere of constant density, of radius $R_0 = cT$. The metric of time and distance shrinks, each by $\alpha = \exp(V/c^2)$, and this study finds the data fitted by $1/\alpha = 1.55$. Earlier, $1/\alpha$ was much larger. As with the deflection of starlight and the Shapiro time delay, gravity affects space like an index of refraction, $n = 1/\alpha^2$. The Hubble concept remains valid, but with geometric distance instead of optical distance. A 2-parameter fit to 3 sets of data finds $T=16.24 \times 10^9$ years and total mass in our big bang is now $M=6.03 \times 10^{52}$ kg (all in our metric). Because n, now 2.41, shrinks with time, all standards of M, L, and T (including atomic standards) are changing by a few parts per billion per decade and should be referenced to a time definite.

## The problem.

Recent data on red shifts (Z) from sources in distant parts of the universe, using SN1a supernovae as "standard candles" to measure distance according to dimness, do not follow a linear Hubble plot of distance vs. Z. From this, it has been inferred that distance increases faster than the red shift. This study uses the data on 19 SN1a events from A. Riess et. al.: (http://journals.uchicago.edu/AJ/journals/issues/v116n3/980111) Two other data sets were analyzed: 42 data points (S. Perlmutter, et. al., AJ, 517: 565-586, 1999 June 1) and 23 data points (B.J. Barris, et al arXiv:astro-ph/0310843 v1 29 Oct 2003). SN1a supernovae are remarkably similar to one another, with brightness of magnitude -19. Many explanations for the non-linearity have been proposed. The leading one calls for an expanding and accelerating universe, somehow powered by dark energy. (Adam Riess, arXiv:astro-ph/0005229 v1 2000)

High Z data to Z=1.2, and some lower Z points, 19 in all, taken from the AJ reference cited above and repeated in Appendix 1, are plotted in Figure 1 as a conventional log-log plot of "magnitude" vs. log Z. The blue line is the Hubble expectation.

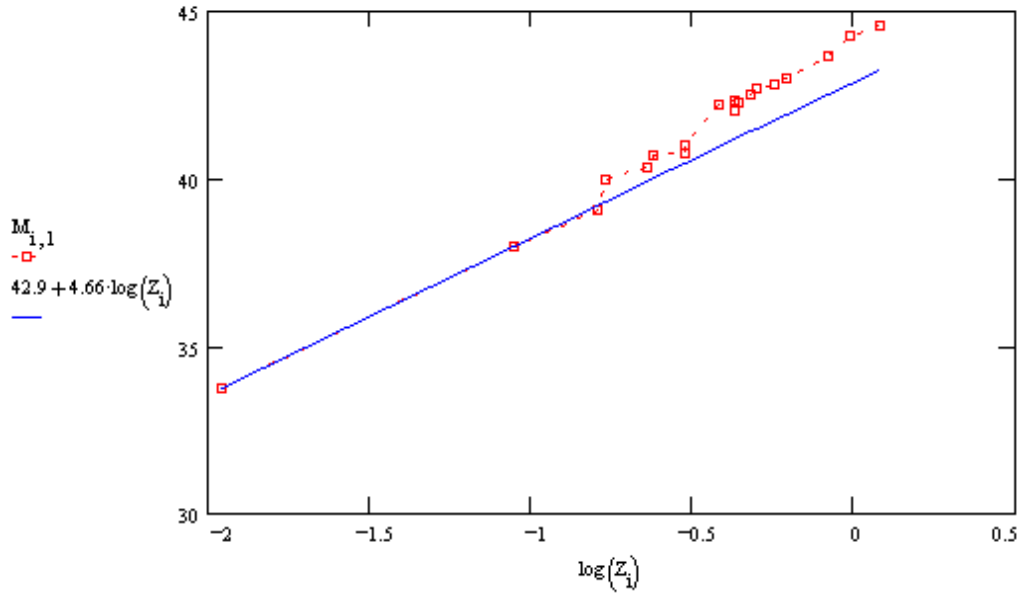

Figure 1. Conventional Hubble log-log plot, magnitude vs. red shift.

# The relativistic metric.

This study finds we can understand the data in terms of relativity. Gravity slows the speed of light, in the metric of an outside observer, and so gravity introduces optical effects into empty space very like an index of refraction. This can be calculated using mass-metric relativity (MMR), a simple alternate to general relativity (GR) which is described in (arXiv.org/abs/physics/0012059 2000). An isotropic reduction of the metric arises from the gravitational potential, V. When gravity is caused by a single large mass, MMR agrees with GR that meter sticks shrink by $\alpha$ where $1/\alpha \equiv 1+GM/Rc^2 = 1-V/c^2$. The gravitational potential V in which we live is much larger than one finds from earth gravity alone, and arises from contributions from all of the mass within the big bang. It is not clear that GR can deal with this many-body problem. That is, since GR is based on SR plus the equivalence principle and assumes gravity arises from a single mass, M, distributed mass poses problems for GR.

MMR is a scalar theory of gravity, and multiple contributions to V are simply additive. In the presence of a gravitational potential, V, whatever its origins, more meter sticks are required to span a given geometric (Euclidean) distance. The effect is identical with the Shapiro time delay, where light behaves as though it were passing through a medium where $n=1/\alpha^2$. The deflection of starlight, too, can then be understood from refraction by applying Snell's law. GR and MMR agree in their accounting for all tests of relativity, so far, but their predictions differ for the gravity probe b. This ongoing experiment may soon decide whether GR or MMR is correct, or maybe something else is needed. More on GR and MMR, in Appendix 2.

GR and MMR each work well in weak gravity, when the metric changes are very small as in our solar system. The gravitational potential in which we live is normally thought to arise mainly from the mass of earth. This overlooks the enormously larger (by $10^9$) gravitational potential arising from all the mass within our big bang. Even though most of this mass is far away, the effect is huge. To see why this is so, recall that the gravitational potential at a point "r" distant from a single point mass, M, is $V = -GM/r$. If we assume the mass density $\rho$ to be constant, throughout our big bang over the large scale, we expect the incremental contribution to V from the mass in a thin shell at distance r from us to be $dV = -G4\pi\rho r^2 dr/r$, and the integral extends from zero to the present big bang radius, $R=cT$. T is the time since the big bang. For a sphere of radius R, filled uniformly with mass, $V = (GM/2R^3)(r^2-3R^2)$. A useful relativistic parameter is $A=GM/Rc^2$, a dimensionless measure of the gravitational potential. Since $A \approx 0.26$, when generally accepted cosmological values are plugged into "A", our present metric is very different from that existing "elsewhere" where there are no gravitational effects and $A=0$. This huge gravitational potential is of little consequence for terrestrial experiments, because it is almost uniform and almost constant.

The value of "A" at sea level, due to the mass of the earth, is only $1.4 \times 10^{-9}$. The Pound-Rebka experiment, on the gravitational red shift of photons, used the Mossbauer effect (sensitive to a part in $10^{15}$) to detect the small change of this parameter over a 10 meter difference of height. When "A" is large, an exponential form for $1/\alpha$, $\exp(-V/c^2)$, is preferable to the linear approximation, $1/\alpha \approx 1+GM/Rc^2=1-V/c^2$. Of course, there is no appreciable difference between these forms for small "A", since $\exp(A) \approx 1+A$ for small A. The large gravitational potential within the big bang makes it possible to distinguish experimentally, from the SN1a data, which of these alternative forms for $1/\alpha$ works better. It is found that the exponential form fits the high Z SN1a data while a linear form does not. From the fit to the data, it will be shown that the metric, of distance and of time, in the place where we live, is now shrunken, relative to "elsewhere", by $e^{-1.5A}=1/1.55$. If we measure distance optically, we find it larger than geometric distance by $n=1/\alpha^2$, i.e. $n=e^{3A}=2.41$, which is the current index of refraction. This index of refraction, larger long ago than now, accounts for the anomalous dimness of the high Z SN1a data. The light had to fight its way through a higher index of refraction, which takes longer.

Understanding the high Z SN1a data requires that we recognize that the metric varies with time. Viewed from "elsewhere", well away from our big bang, our metric of optically determined distance is shrunken by an index of refraction, $n=e^{3A}=2.41$. The time since the big bang, as we measure it, is 16.24 billion years. Measured by a observer "elsewhere", it is only 1/2.41 of this, 6.76 billion years. This means the Shapiro time delay, for very distant SN1a data, can be many billions of years! Further, our metric was still further shrunken, long ago, when the big bang was smaller and the gravitational potential was greater. The SN1a signals were sent when our big bang was smaller, and the information encoded in those signals reflects the larger gravitational potential. With meter sticks shrunken even more than at present, the optical distance to the high Z SN1a's we have considered is many times larger than the geometric distance.   Conventional Hubble plots assume measured distance is geometric distance .

## Correcting the data for velocity and common time.

We need to put the data from Figure 1 into better form for analysis. The Hubble concept is that the present velocity of recession is proportional to the present distance, and no change of metric is contemplated.  The velocity, $\beta c$, can be  found by SR from Z.  The red shift parameter Z is defined by the increase of wavelength, $\lambda$:

$$Z = \delta\lambda/\lambda = (\lambda' - \lambda)/\lambda \qquad [1]$$

When velocity is small relative to c,  $Z \approx \beta$.  At higher recession velocities, SR is needed to extract $\beta$:

$$Z=(1+\beta)/[(1-\beta^2)^{1/2}] - 1 = [(1+\beta)/(1-\beta)]^{1/2} -1 \qquad [2]$$

$$\beta=[(Z+1)^2-1]/[(Z+1)^2+1] \qquad [3]$$

We must also change the plot to one of *present* distance vs. velocity, in the spirit of the Hubble concept, since the SN1a's blew at various times up to many billions of years ago.  The larger the red shift, the earlier the time when the event happened. To find a distance appropriate to a Hubble plot, one must extrapolate observed distances to where, absent acceleration, they are at this time.

We don't yet know the place of origin of the big bang.  Any place within, other than at the periphery and we are certainly not there, appears to be the source of the expansion since everything is observed to be departing that place.  Following the big bang, at t=0, the remnant which contains the SN1a that will ultimately be observed moves away from the observer at velocity $\beta c$.  At time t=T* when the SN1a blows up, the observer's geometrical distance from it is "r". At a later time t=T, when the light is actually seen by us, the observer's distance from the source has increased to r'=r(1+$\beta$) because of the time it takes the light to travel to us.  In the spirit of Hubble's insight, absent any acceleration, one expects that all values of r, extrapolated to  constant time, will be proportional to their recession velocity, $\beta c$, found using [3].  All  material within the big bang is enclosed within a sphere of radius $R_0$ =cT.

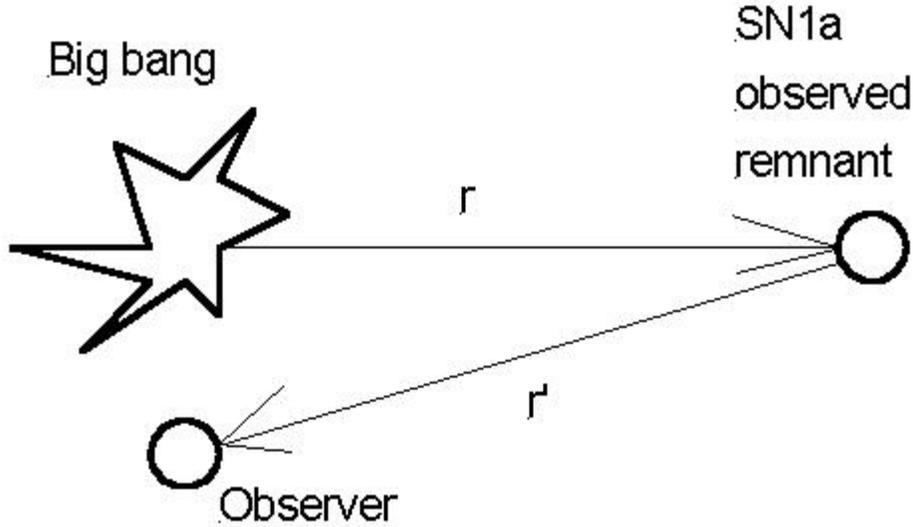

Figure 2. Assumed geometry of the big bang.

Absent acceleration, the remnant containing the SN1a will have moved at constant velocity, $\beta c$, relative to the observer, for a time $T^* = r/\beta c$, and it then emits a light signal. This light signal encodes both the distance (by the dimness of the "standard candle") and the velocity of recession (by the red shift, Z), as they were at $T^*$. The light signal then traverses the distance r' to the observer, i.e. $r(1+\beta)$, in the additional time, $\tau = (1+\beta)r/c$. Notice that r' is the distance measured by the astronomer, by the dimness of the event. The total elapsed time, T, since the big bang, is a constant for the earth-bound observer.

$$T = T^* + \tau = r(1/\beta c + (\beta+1)/c) = (r/c)((1+\beta+\beta^2)/\beta) \qquad [4]$$

Since T is independent of r, r is $\beta cT/(1+\beta+\beta^2)$. That is, r alone is not proportional to $\beta$, but its extrapolated value, $r(1+\beta+\beta^2)$, is. The effect is not large, save at high Z. To extrapolate "r" to the present value (i.e. where the SN1a residue is at time T), r must be increased by $T/T^*$. This ratio is available from [4]. Since T is proportional to $1/\beta + 1+\beta$, and $T^*$ is proportional to $1/\beta$:

$$T/T^* = (1/\beta + 1+\beta)/(1/\beta) = 1+\beta+\beta^2 \qquad [5]$$

The extrapolated (to time T) distance R is found by first calculating r from the observed r′, $r = r'/(1+\beta)$, and then augment it by $T/T^*$. At time T, R is:

$$R = [r'/(1+\beta)][1+\beta+\beta^2] \qquad [6]$$

It should be noted that this correction, alone, does not explain the curvature of the Hubble data, shown in Figure 1, and in fact it increases the curvature.

We can either correct r′ to R, as in [6], and plot R vs. β, or we can plot r′ vs. β(1+β)]/[1+β+β²]. We choose to do the latter, so that r′ remains the observed distance. In the spirit of the Hubble concept, one then expects a linear plot. Any deviation from linearity indicates complications. But with these efforts to convert red shifts into velocities and to extrapolate distances to a common time, the non-linearity of the data plot only increases as is shown next.

$$R_i := \left[ \beta_i \cdot \frac{1+\beta_i}{1+\beta_i + (\beta_i)^2} \right] \quad R_i := R_i \cdot 15.3 \cdot 10^9$$

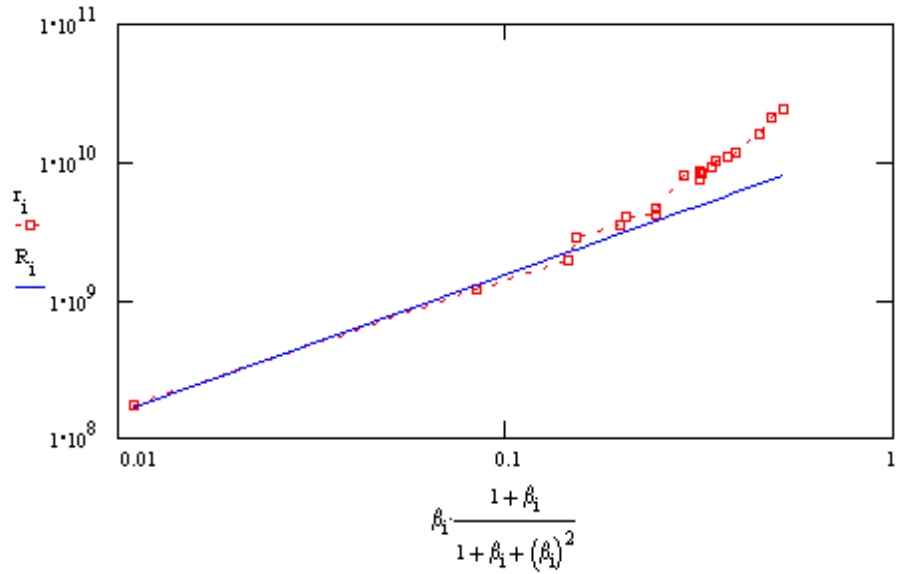

Figure 3. Hubble log-log plot, Z corrected by SR and common time, T.

A log-log plot crowds together the data at high Z and large distance. To see more clearly the non-linearity at high Z, the data is next plotted linearly in Figure 4. The deviation from linearity is even more apparent! Notice that the maximum value possible for the abscissa, when β=1, is 2/3.

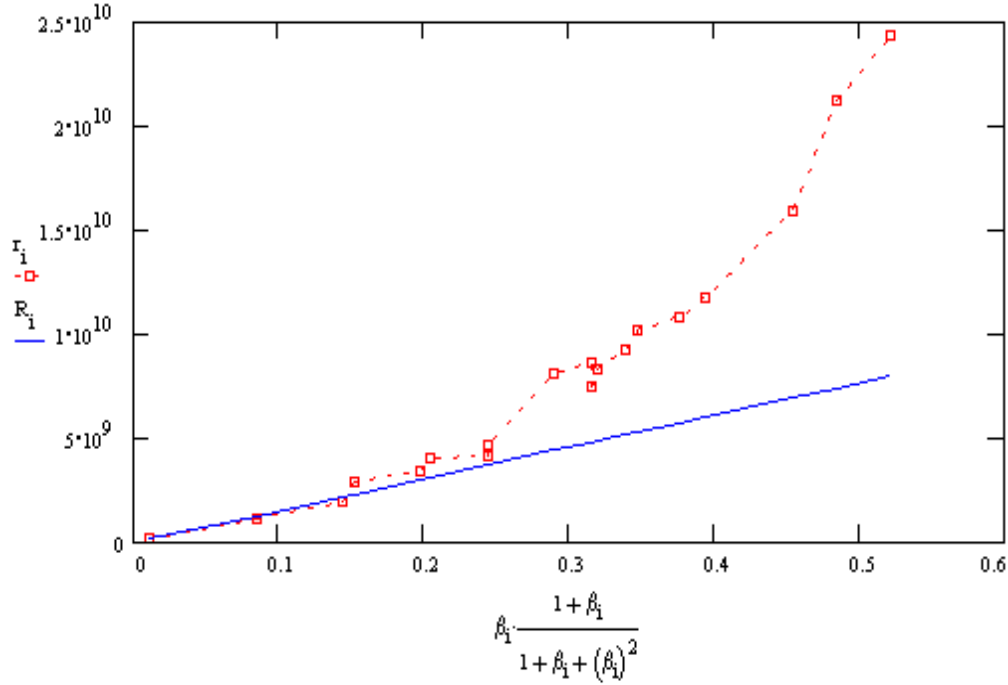

Figure 4.  Corrected data plotted linearly.  Blue line is H=59.3.

## Optical distance exceeds the geometric distance.

Figure 4 is the starting point for the relativistic analysis, with a linear plot of observed distance vs. β as corrected for Z and for common time. The low Z data is little affected by the change from Z to $\beta(1+\beta)/(1+\beta+\beta^2)$.

As mentioned earlier, the reduction of the metric in a gravitational potential V is equivalent to imbuing space with an index of refraction, $n=1/\alpha^2$, where $1/\alpha \equiv 1+GM/rc^2$ for a single point mass, M, and V = -GM/r. This provides a simple explanation for the Shapiro time delay of light passing near the sun, and for the deflection of starlight during a solar eclipse. When the effect is large, we consider an alternate form of $1/\alpha$, $\exp(-V/c^2)$, and find it works better. With an index of refraction larger than "1", the optical path length exceeds the geometric distance, R. This makes the SN1a light signatures appear dimmer than they would be, were n=1.  This would not, in itself, lead to curvature if "n" were constant.   But "n" was larger, earlier, and this increases the differential optical distance along the light path.

In the early universe, the radius $R_0$ of the big bang was smaller and the mass was bigger.  The gravitational potential was then larger.  Consider the present gravitational potential within a sphere circumscribing our big bang.  On the large scale, assume a constant mass density, $\rho$.  The total mass of the big bang is presently M.  The gravitational potential within a sphere containing distributed mass is more complicated than simply V= -GM/r, but can be calculated from the sum of all M/r terms, and takes the form shown next for $r<R_0$:

$$V = (GM/2R_0^3)[r^2 - 3R_0^2] \qquad [7]$$

This is plotted in Figure 5.  The uppermost line is for now, T, and the abscissa is $r/R_0$.  Successive lines are for earlier times, T*, back to T/2.  For $r>R_0$, V=-GM/r at all distances.  All distances are geometric distances.

Remnants of the big bang, even those moving at near c, can only reach as far as $R_0/3$ before emitting any signal that can possibly reach us now.  That is, the outward bound remnant must explode before T/3, since it will take the light signal 2T/3 to return because we continue to recede from the remnant at c.  Slower remnants can wait longer, even up to T at very slow recession velocities, since it doesn't take as much time for the light signal to reach us.  No matter how high the red shift, we can only look back to T/3.

In the analyses that follow, the present total mass of the big bang is M.  MMR relativity holds that rest mass increases with gravitational potential by $1/\alpha \equiv 1+GM/Rc^2 = 1-V/c^2$.  The length of a meter stick shrinks, by $\alpha$.  In the curve fitting process, we calculate the "optical" distance by integrating, over the light path, ncdt.  The gravitational potential, from which one finds "n", has two terms. To simplify the calculations, R replaces $R_0$. At center, the gravitational potential term is presently $V=-1.5GM/Rc^2$. The index of refraction is  $n=1/\alpha^2$.  Earlier, at T*, M* was greater and R* was smaller.  We correct $1.5GM/Rc^2$ to an earlier time, t, by multiplying it by $T^2/t^2$ and then square the exponential form of "$1/\alpha$".  The time dependence of A is $A=GM/Rc^2(T^2/t^2)$, and this corrects for temporal changes of  mass and distance.

We will fit the SN1a data to two models.  The simple model omits any effects due to the "$r^2$" term in [7]. The "better model"  adds the "$r^2$" term in [7], both as it was at the time the SN1a blew and as it changes during the time it took for the signal to reach us from the far reaches of space.

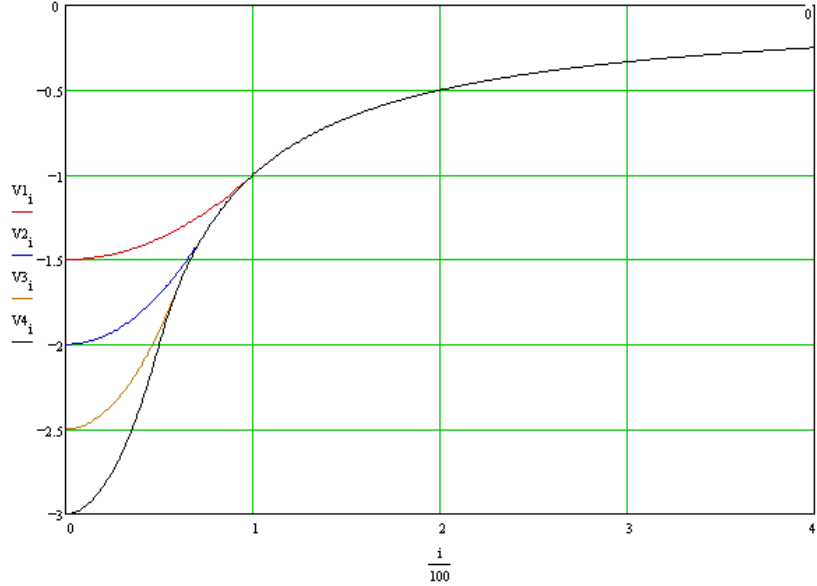

Figure 5. Gravitational potential for sphere, at different times.

Suppose that, sometime before our big bang, there accrued an immense aggregation of mass. The gravitational potential at the center of this compact proto-BB was large and negative and extended towards -∞ in Figure 5. The metric of space was highly contracted in the vicinity of this mass, by the huge gravitational potential. Comes then the big bang, about 16.24 billion years ago, according to our present-day clocks, with a release of energy which overwhelmed the gravitational attraction which had assembled all this mass. Fragments of the mass contained in the proto-BB then moved outwards, at different speeds, populating space uniformly, and the metric inflated in accordance with the falling gravitational potential. The potential at the center (r=0) rose towards its present value at t=T, which is V1. At an earlier time, T/2, V was twice as large (V4). At time t=T*, where T<t<T/3, each SN1a in the data set blew up and sent its message to us.

Since we are looking back in time, each SN1a blew in a gravitational potential larger, by $T^2/T^{*2}$, than that which we now experience. Total mass M was larger, and R was smaller. By [7], the gravitational potential at distance "r" from center is, and was,

$$V = -3GM/2R_0 + 0.5(GM/R_0^3)(r^2) \quad \text{at T} \qquad [8]$$

$$V^* = -3GM^*/2R_0^* + 0.5(GM^*/R_0^{*3})(r^{*2}) \quad \text{at T*} \qquad [9]$$

## Curve fitting to 19 data points: a simple model.

The 19 data points are listed in Appendix 1. The internal precision of this data is remarkable. Where are we, within the big bang? Where is the center of the big bang? We really don't know. We simplify the analysis by assuming that we are near the center, as in Figure 2. With better data, we may be able to do better.

In the simple model, we completely neglect the "$r^2$" terms in [8] and [9]. That is, we take it that the index of refraction is the same, everywhere within the big bang, and is the gravitational potential at r=0. Since the SN1a data arose at different times, T*<T, the index of refraction of space was larger then than now. The simple model recognizes changes, with time but not with distance, of this index of refraction as the light signal moves to us.

The data was analyzed for two different mathematical forms, $1/\alpha = 1+1.5GM/R_0c^2$ and also $1/\alpha = \exp(1.5GM/R_0c^2)$. A much better fit was found with the exponential form of $1/\alpha$. Notice that $1/\alpha^2 = \exp(-2V_0/c^2)$

$$n = 1/\alpha^2 = \exp(-2V_0/c^2) \qquad [10]$$

$$n^* = 1/\alpha^{*2} = \exp(-2V_0^*/c^2) \qquad [11]$$

Let $A=GM/Rc^2$ for calculation. It was larger, earlier, because "M" was larger and "R" was smaller, each by a $1/\alpha$ factor. Hence,

$$A(t) = A(T/t)^2 \qquad [12]$$

Since $T^* = T/(1+\beta+\beta^2)$, when each SN1a blew, the optical distance is found by integrating "cndt" from the time each SN1a blew, $t=T/(1+\beta+\beta^2)$, to t=T.

$$n = 1/\alpha^2 = \exp(3A(T/t)^2) \qquad [13]$$

The index of refraction, n*, larger at earlier times, is shown for each data point in the next graph, Figure 6. Notice that T* can only range, at most, from T/3 to T, as mentioned above. The available data set begins at T*/T = 0.48 (Z=1.2) and extends nearly to 1.0. The calculation below uses the parameters A and T which best fit the better model that includes the $r^2$ term.

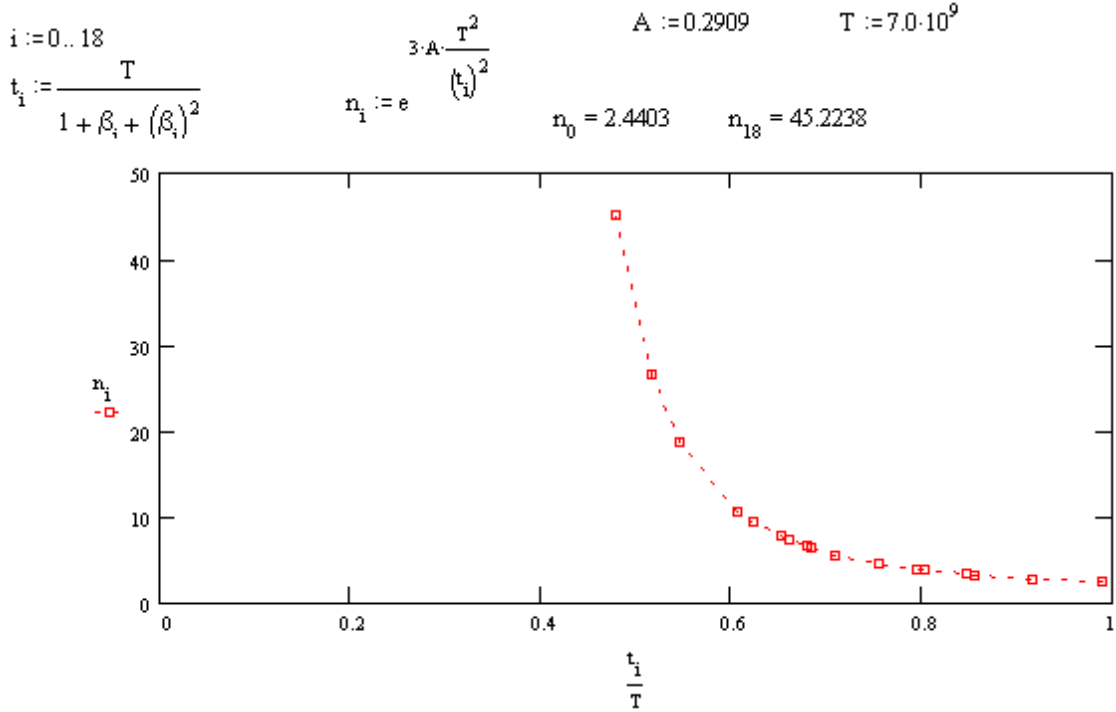

Figure 6. Index of refraction of space, at the time each SN1a blew.

The optical distance, D, is the integral of "cndt" over time since the SN1a blew, $T/(1+\beta+\beta^2)$, to T. This is equivalent to r′ in [4], i.e. the optical distance from the SN1a to us. The parameters A and T were varied to find the best fit to [14]. The value of "c" is 1, in units of light years/year.

$$D_i = c \int \exp[3A(T/t)^2] \, dt \qquad [14]$$

Notice that the values for "n" in Figure 6 for the nearest and the farthest data points are shown, at the time when the SN1a's exploded, and are 2.44 and 45.2. The curve fit to the simple model is shown next, where SD is the standard deviation and is $7.863 \times 10^8$ ly for best fit.

$$D_i := \int_{\frac{T}{1+\beta_i+(\beta_i)^2}}^{T} e^{3 \cdot A \cdot \left(\frac{T}{t}\right)^2} dt \qquad T := 8.9 \cdot 10^9 \quad A := 0.234$$

$$S := \sum_i (r_i - D_i)^2 \qquad SD := \sqrt{\frac{S}{18}}$$

$$SD = 7.863 \cdot 10^8$$

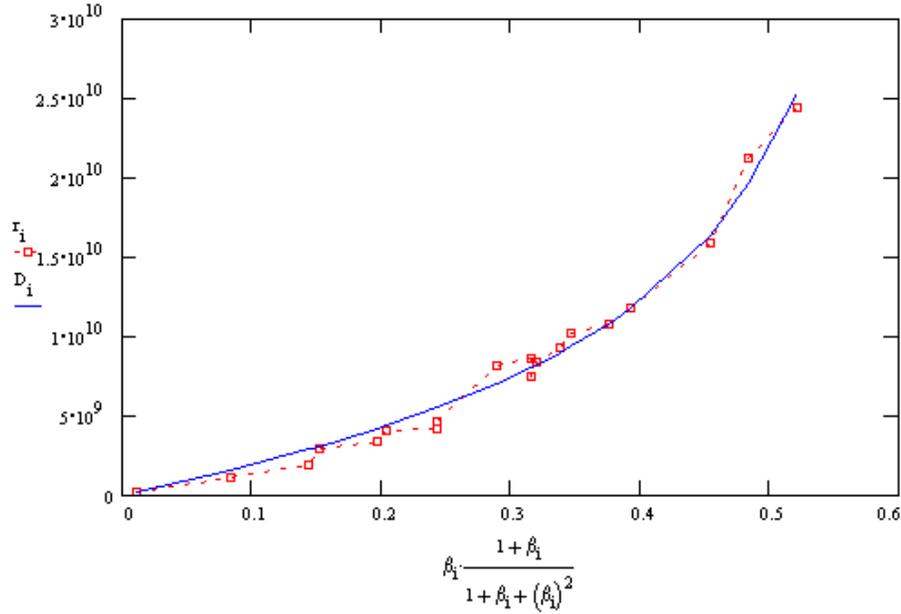

Figure 7. Curve fit of SN1a data to simple model, H=54.5.

The data points, shown as red boxes, rise above a linear Hubble expectation (not shown), because of the larger "n" encountered early during each one's long journey. The blue line is theory, [14].

## Curve fitting: a "better" model.

A better model is displayed in Figure 8, and includes the second (quadratic) term in the potential. Potentials existing within the big bang are shown for an earlier time, $t=T^*$, and for the present time, $t=T$. The edge of the big bang is shown as $r=1$ now, and was only "½" at $t=T^*=T/2$. The potentials at $r=0$ are -1.5 at present and -3.0 at $t=T/2$. We improve the simple model by recognizing that each SN1a moves outward a distance r before blowing up. The gravitational potential in which each SN1a explodes is then smaller than in the simple model, according to its distance from center. The light signal moves through space to us (shown as $r=0$ and $V=-1.5$ in the graph). Slow remnants will move a short distance r before exploding, while fast-moving remnants will move a larger distance r. We want to account for these changes of potential, both at the time of the explosion and also during the light path to us. The diagram may mislead one into thinking that the length of the light path is equally long in both cases. The "r, low β" horizontal line takes a long time before it blows, during which the potential rises to near our present potential where we receive the signal. That light path is hence much shorter than that from a fast SN1a.

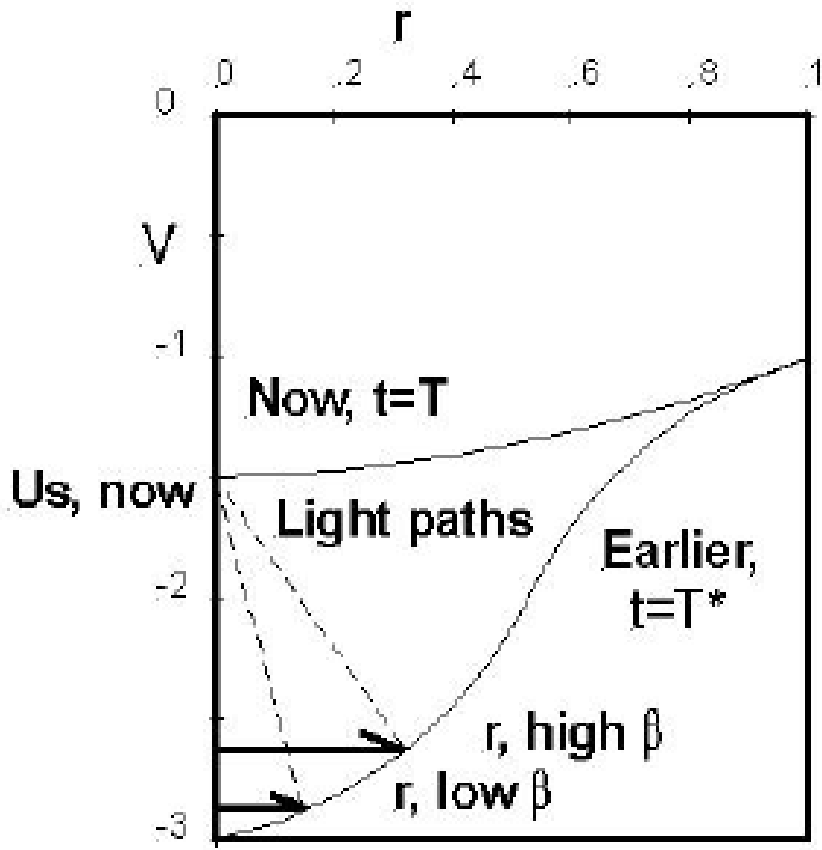

Figure 8. A better model, including the "$r^2$" term.

The index of refraction for this better model is:

$$n = \exp(3A(T/t)^2 - A(T/t)^2\beta^2[(T/t-1)/(\beta(1+\beta))] \qquad [15]$$

The $(T/t)^2$ term corrects A, which is the potential at r=0, according to the mass M and the radius $R_0$ of the big bang at time t. $\beta$ is the ratio of $r/R_0$. That is, $r=\beta ct$ and $R_0=ct$. The term in square brackets accounts for shrinkage in the "$r^2$" term as the light moves towards the observer. By [5], $T/T^* = 1+\beta+\beta^2$, and the shrinkage factor is:

$$[(T/t-1)/(T/T^*-1)] = [(T/t-1)/\beta(1+\beta)] \qquad [16]$$

$$D_i = c \int \exp\{3A(T/t)^2 - A(T/t)^2\beta_i^2[(T/t-1)/(\beta_i(1+\beta_i))]\}\, dt \qquad [17]$$

As before, the integral is from $t=T/(1+\beta+\beta^2)$ to $t=T$. This model obtains a better fit than that found using the simple model, SD=$7.3046 \times 10^8$.

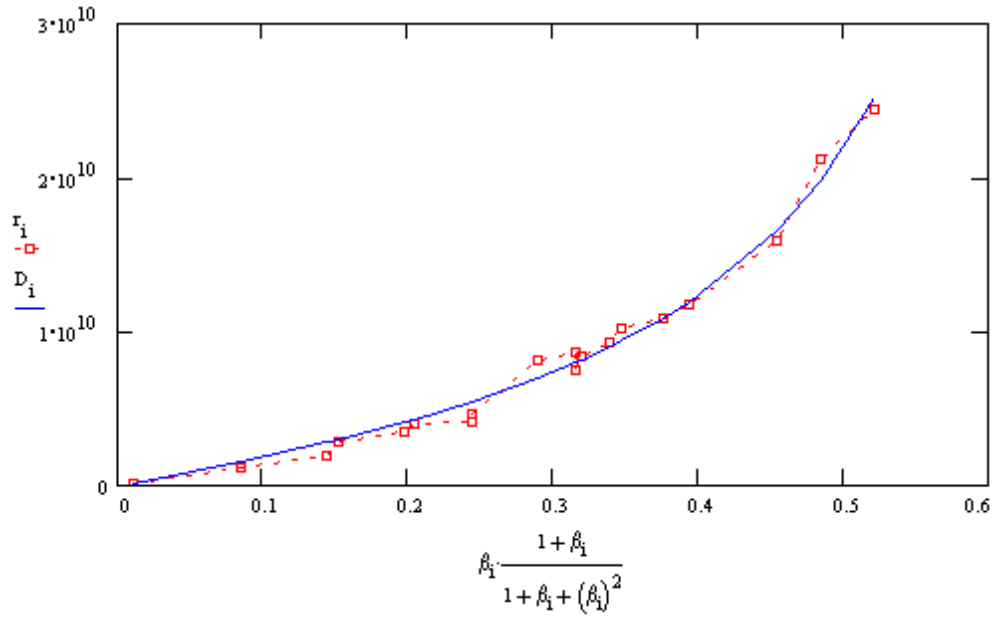

Figure 9. Curve fit to better model, H=58.4.

## Curve fitting to another data set, 42 data points, simple model.

This set of 42 data points is from the work of Perlmutter and his many collaborators. (S. Perlmutter, et. al., AJ, 517: 565-586, 1999 June 1) It is obvious from the data, plotted below, that the internal precision is not as good as the set of 19 data points, but it is interesting to analyze this different data set and find out how well the fitted parameters agree.

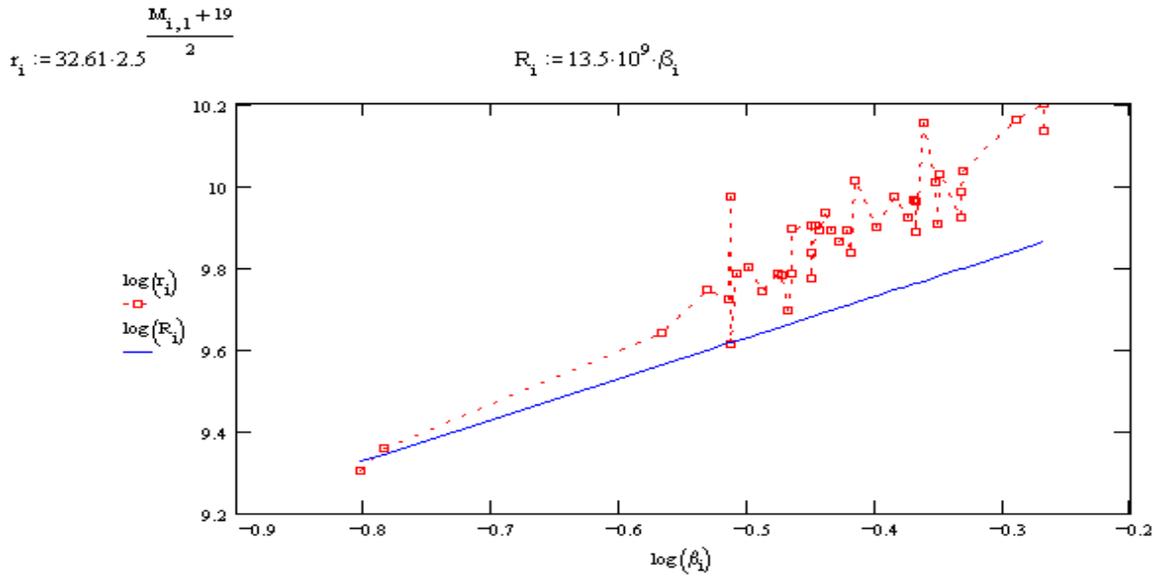

Figure 10. The 42-point data set, log-log plot of r vs. β.

The blue line is the Hubble expectation, i.e. distance is proportional to recession velocity. The recession velocities do not cover as large a range as in the previous data set, and so the blue line has been added to show the Hubble expectation. The same line is added to the linear plot, below; the abscissa has been corrected for constant time, T. In this linear plot, the blue line of course passes through the origin, 0,0.

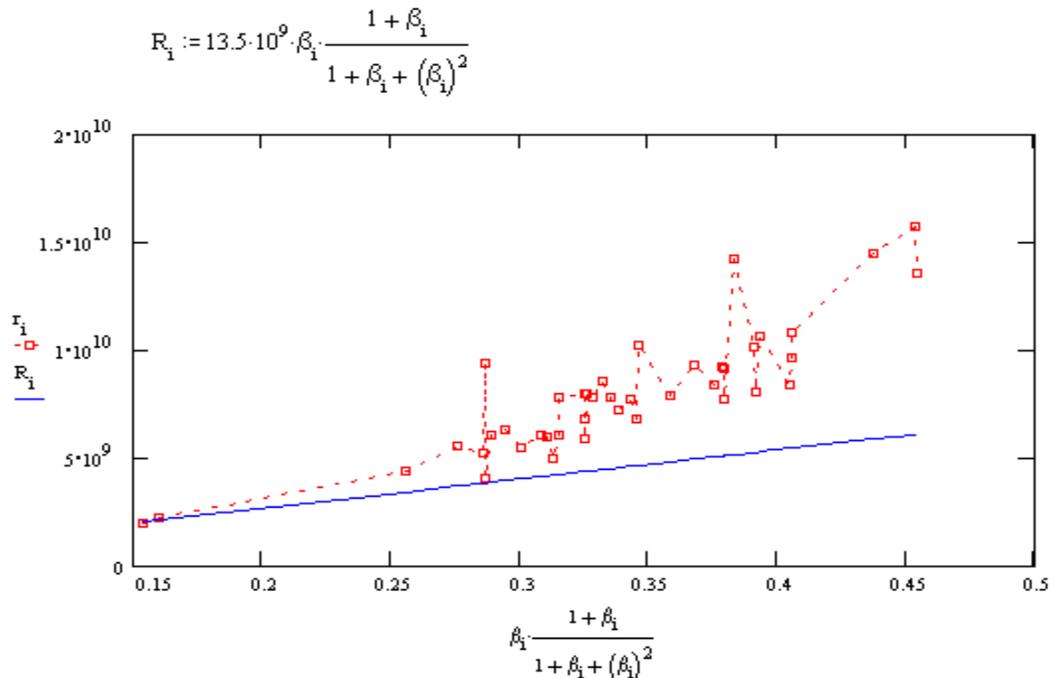

Figure 11. The 42-point data set plotted linearly, corrected for constant T.

The curve fits for this data set, to the simple and the better models, are shown next.

$$D_i := \int_T^T \frac{e^{3 \cdot A \cdot \left(\frac{T}{t}\right)^2}}{\sqrt{1 + \beta_i + (\beta_i)^2}} \, dt$$

$T := 6.5 \cdot 10^9 \qquad A := 0.264$

$$S := \sum_i (r_i - D_i)^2 \qquad SD := \sqrt{\frac{S}{18}}$$

$SD = 2.0543 \cdot 10^9$

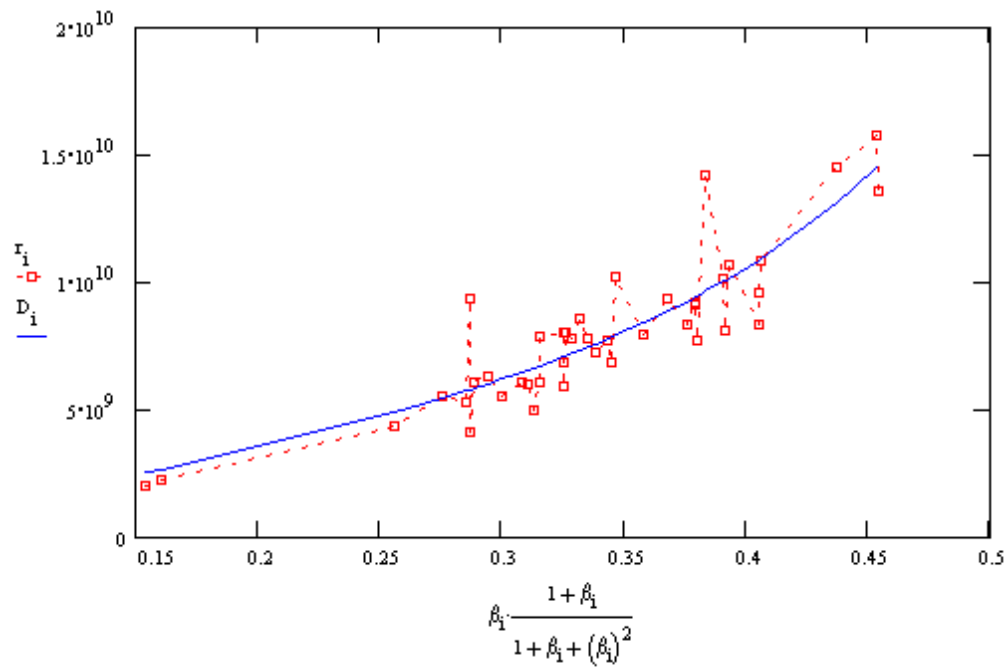

Figure 12. Data fitted to simple model.

$$D_i := \int_T^T \frac{e^{3 \cdot A \cdot \frac{T^2}{t^2} - A \cdot \frac{T^2 \cdot \beta_i}{t^2} \cdot \left[\frac{\left(\frac{T}{t}\right) - 1}{1 + (\beta_i)}\right]}}{\sqrt{1 + \beta_i + (\beta_i)^2}} \, dt$$

$T := 5.3 \cdot 10^9 \qquad A := 0.3142$

$$S := \sum_i (r_i - D_i)^2 \qquad SD := \sqrt{\frac{S}{18}}$$

$SD = 2.0565 \cdot 10^9$

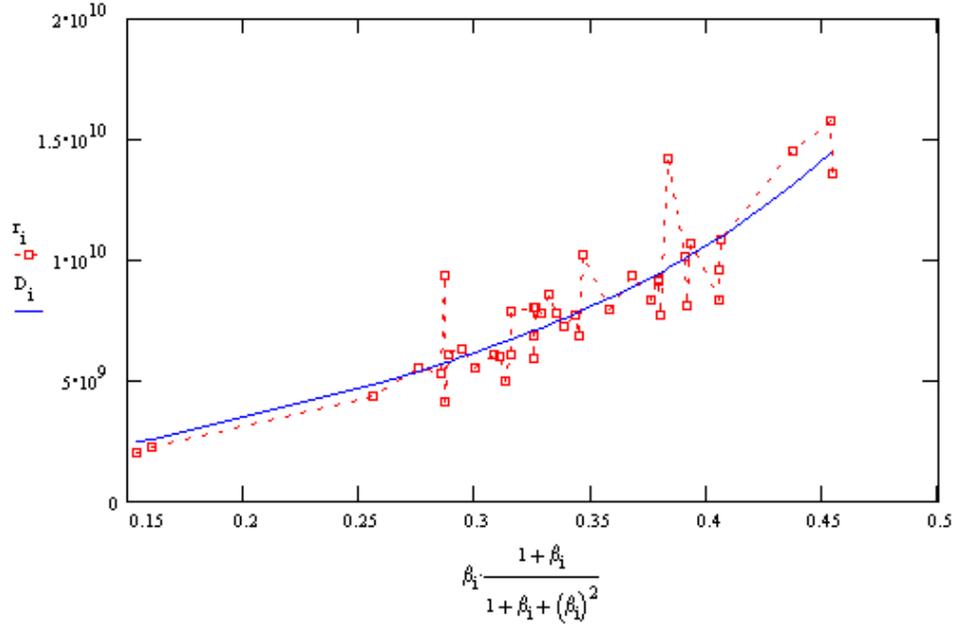

Figure 12. Data fitted to better model.

A third set, of 23 data points, (B.J. Barris, et al arXiv:astro-ph/0310843 v1 29 Oct 2003) was also analyzed . Again, the scatter was larger than in the 19 point data set. The fitted parameters for A and T are shown in the next table, which also lists the derived information.

|  | 19 point data set | 42 point data set | 23 point data set | Weighted average |
|---|---|---|---|---|
| Model: | Better | Better | Better |  |
| $10^{-9}$ T years | **7.0** | **5.3** | **6.4** | **6.76** |
| A | 0.2909 | 0.3142 | 0.278 | .293 |
| $1/\alpha = e^{1.5A}$ | 1.5471 | 1.6021 | 1.538 | 1.551 |
| $n = 1/\alpha^2$ | 2.3934 | 2.5666 | 2.366 | 2.406 |
| $10^{-52}$ M kg. | 6.1886 | 5.4274 | 5.509 | 6.028 |
| $10^{30}$ ρ, g/cm³ | 3.7029 | 6.0665 | 4.467 | 4.031 |
| $10^{-9}$ R **lt yr** | 16.754 | 13.603 | 15.14 | 16.242 |
| H | 58.4115 | 71.9388 | 64.64 | 60.525 |
| $10^{-8}$ st. dev. | **7.3046** | 20.565 | 18.12 |  |

**Table 1. Comparison, fitted parameters and derived quantities.**

## Discussion

This study was undertaken in an attempt to understand the reason for the non-linearity found in Hubble plots of recent SN1a data from deep space. A relativistic change of metric according to the gravitational potential has been found to fully account for the non-linearity. When this explanation is accepted, our view of the big bang will have changed markedly. Instead of a universe with constant mass, expanding maybe forever and maybe not, we will have accepted that the very metric in which we live is controlled by all the mass within the big bang. Our fundamental standards of mass, length, and time will have been shown to be time-dependent, according to the present size of the big bang.

Using the weighted average of the parameters for the 3 sets of data, the time since the big bang is 6.76 billion years in the metric of an observer away from all gravity. This observer recognizes that locals will use light signals to measure distance, which increases the radius of the big bang, by "n", from 6.76 billion light years to 16.24 billion light years. The Hubble constant is 60.5, as we locals reckon it. The use of a clock, in earth metric where clock rate varies according the total mass and radius of the big bang, is fraught with difficulties. The "elsewhere" clock is inherently linear, but our clocks run slower as time passes. Long ago, our clocks ran a lot faster than presently. For this reason, attempts at radiological dating should incorporate the changing metric of time. See Appendix 3.

Is there any compelling evidence to support this theory? Yes. This relativistic model for the SN1a data provides an <u>independent</u> measure of the total gravitational mass (including dark matter) within our big bang, $6.03 \times 10^{52}$ kg, quite close to the current estimate of $6 \times 10^{52}$ kg. http://www.astronomynotes.com/cosmology. So, the MMR concept both explains curvature in the SN1a data and independently finds a total mass similar to that found by other means.

If the mass were constant, one could predict from Newtonian gravity whether the big bang is open or closed. But, the present study finds that the total mass changes with time. More mass, long ago. Less mass, in future. In the presence of other masses, a given mass is larger than it would be elsewhere away from gravity. The present mass density is $4.03 \times 10^{-30}$ g/cm$^3$. This is already less than the mass density presently thought necessary for closure, (http://www.curious.astro.cornell.edu), $3H^2/8\pi G = 10.6 \times 10^{-30}$ g/cm$^3$.

The length (and breadth, and thickness) of a meter stick shrinks in the presence of gravitational potential, by a factor currently 1/1.55. As the big bang grows, and assuming it is "open", meter sticks are becoming larger than at present.

The cosmological principle holds that, on a large scale, all is the same, in all directions, no matter how far one goes. This study suggests the presence of a measurable quantity, absolute gravitational potential, that varies from place to place within the big bang. That is, there is a structure at the highest level. Maximum at the center, the gravitational potential falls off, outwards, by 1/3. If this be true, we may hope to eventually find out where we are and where is the center of the big bang.

If the use of MMR to calculate the gravitational metric consequences is off-putting to the reader, he may wish to try to calculate the increase of measured distance under gravity using GR. The result should be the same. GR and MMR are equally able to explain the observed metric consequences when gravity arises from a single concentrated entity, as in the Shapiro time delay for light passing near the sun on its way to and from a transponder on Mars. Or, for the deflection of starlight during a solar eclipse. Although GR finds that gravity affects the radial component of the metric according to $1+2GM/Rc^2$, it does not interpret this in terms of a gravitational potential! It is not clear that GR can deal with the metric consequences arising from multiple masses. GR is a vector-tensor theory of gravity and focuses on the equivalence of gravity and acceleration, where acceleration is towards a single mass. Gravitational potential is scalar and arises from many sources within the big bang. Acceleration need not be a factor in gravitational potential. MMR is a scalar theory of gravity, and multiple contributions to the gravitational potential are simply additive. No matter how one calculates it, he must find that the number of meter sticks needed to span a given geometric space increases when the gravitational potential along the path is larger than that at the observer.

The cosmological metric found in this study is truly surprising. We tend to think that n=1 in outer space in the solar system. If "n" is really so large, n=2.41 at this time, it means that our terrestrial metric is dominated by the gravitational potential of all the mass in our big bang. This "n" was even larger, earlier, and will decrease to n=1 (if the big bang is "open").

It is disturbing to recognize that these parameters change with time. "n" was larger, earlier. The total mass of the big bang was then greater, and the radius was smaller. This means that the mass density was very much larger than that existing now. Conventional cosmology assumes that the total mass within our big bang is constant, and calculates the likelihood of a closed or an open universe based on the gravitational tug of the receding masses on one another. Even if, under the assumption of constant mass, it were found that sufficient mass now barely exists for eventual closure, the thrust of this study is to say closure still will not happen. That is, the more widely distributed is the matter, the less becomes its mass. And so, less and less gravitational force is available for closure. The assumption that mass density is uniform, on a large scale, is generally supported by the fit of the data to the "better" model. Examined closely, however, the fit is not quite right for low Z data. The fitted line is a bit higher than the data, which suggests that the mass density may be less in our vicinity than exists at larger distances.

The high-Z SN1a data probes the gravitational potential within the big bang, both here and now and also at other places at earlier times. It provides an independent measure of the radius of the big bang and of the total mass of all the particles in the big bang; from these one can calculate the density of matter. The relativistic explanation, however, is accompanied by its own cosmological problems. We have to accept that our metric is not the immutable "clean" standard we have long assumed. The search for such immutable standards has led to the use of atomic masses and time intervals, replacing physical measuring rods and clocks. But all standards, even these atomic standards, are changing by a few parts per billion per decade. In the search for a solid rock on which to found a metric, we are forced to consider a hypothetical "elsewhere" where gravitational effects are absent. Only there is rest mass a minimum, meter sticks have maximum length, clocks run most slowly, and nothing changes with time. Such a place does not exist within our big bang. The gravitational potential where we live, now large, decreases with time. Our analysis of events within our big bang must recognize conditions as they existed at earlier times, and it appears that the metric existing then differs from that now. Since even our atomic standards of mass, length, and time are slowly changing in consequence of the expansion of our big bang, it might be appropriate to tie these standards to a definite time such as 12:01 GMT 1/1/2000.

This finding of a changing metric within our big bang requires that we rethink the basics of cosmology. For example, our most firmly established laws of physics include conservation of linear momentum and conservation of angular momentum. Consider a remnant of the big bang. Hubble found that the velocity of recession is proportional to distance from us. Cosmology holds that this velocity should slow, over time, due to the gravitational pull of other masses. But if linear momentum, $p=mv$, is truly constant, then the decreasing mass as the big bang proceeds could cause the velocity to increase! These two accelerations are in opposite directions, and they may cancel. Or, if the outward acceleration exceeds the gravitational acceleration, one might surmise that "dark energy" or something like that is acting like an anti-gravity force. There are many other things to consider, such as the consequences of the changing metric on the evolution of life. The solar day is an inverse measure of angular velocity and has been assumed constant except for lunar distance change and small net losses of atmosphere to space or gains due to meteorites. In the distant past, when mass was larger and distances were shorter, the solar day was shorter. With the better data coming from the SNAP space telescope, we can more accurately account for past and future changes of the metric.

Many cosmologists refuse to recognize that red shifts arise from Doppler motion, preferring instead to assign them to inflation as the metric changes. This is denied by MMR, since any change of wavelength by inflation is mirrored by the change in the length of the meter sticks we use to measure the wavelength.

Whether gravitational potential is absolute or relative is debatable. When gravitational force is derived from the gradient of the potential, any constant may be added to the potential and this does not change the force. The thrust of this study is that there is an absolute gravitational potential, which increases mass and so changes the metric in ways that can be measured. In MMR, we are not free to add an arbitrary constant. The arbitrary setting of the zero of gravitational potential, as we normally do when calculating gravitational force on earth, is a simplification useful only because the much larger gravitational potential due to the big bang is almost constant. Recall the flap about the magnetic vector potential, previously thought to be only a mathematical crutch for the calculation of magnetic fields. Its independent existence was doubted until the suggestions by Aharanov and Bohm which led to experimental evidence of its existence.

There may be a problem with using a SN1a as a standard candle, when it ignites in a different gravitational metric. For example, at half the present lifetime of the big bang as seen from elsewhere, all masses were twice their present value and all distances were half their present value. A SN1a event has been determined to require a definite number of hydrogen atoms, several times those present in our sun, in our current metric. If the mass density were really 16 times greater, it seems likely that fewer hydrogen atoms would be needed for ignition. It seems reasonable that fewer hydrogen atoms means that fewer photons are produced when the SN1a ignites, and so these distant events may be intrinsically dimmer than from nearby SN1a's. Such matters are best dealt with by experts in fusion. This does not deny the premise of this study, that relativistic optical density can explain the "dimmer than expected" data at high Z. It merely suggests that part of the answer may lie in a smaller intrinsic brightness of a SN1a when it ignites in a greater gravitational potential.

## Summary

The non-linearity of Hubble plots of high Z SN1a data has been troubling. To many, it meant that recession speed is increasing instead of being slowed by gravity. Speculations include the possibility of dark energy leading to repulsion instead of gravitational attraction, thoughts about Einstein's cosmological constant, and others. These speculations are so persuasive that the new space telescope designed to get more and better SN1a data is named SNAP, standing for supernova acceleration project. This study finds no evidence for acceleration, and finds that the "problem" can be understood by recognizing the relativistic changes in the metric. The approach taken here is to first recognize that the Hubble concept holds that, absent acceleration, the present recession velocity is proportional to the present distance. The recent SN1a data is from events that occurred long ago, and the remnants of those events are no longer at the place where they were when each event took place. To plot these in accordance with the Hubble concept, we must first make the necessary corrections for a common time, and also convert red shift Z to velocity; but we find that this only increases the non-linearity in Hubble plots. After making these corrections, it has been found possible to account fully for the non-linearity using relativistic corrections to the metric based on differences between the gravitational potentials now and those existing at the SN1a's when they exploded. The metric consequences of gravity include that we recognize that our standards of mass, length, and time differ from those "elsewhere" where there is no gravitational potential. A 2-parameter fit is obtained for the available high Z data. The net result confirms the Hubble concept, that recession velocity is proportional to geometric distance and is valid over billions of years and distances of billions of light-years. This absence of acceleration implies an open universe. Interpreted in this way, using the weighted averages of the fitted parameters for 3 sets of SN1a data, one finds $A=GM/c^2R_0=0.293$ and $R_0=16.24 \times 10^9$ light years in our metric, and so the total M of the big bang is presently $6.03 \times 10^{52}$ kg. The present Hubble constant for best fit is 60.5, again in our metric. If the metric explanation for the SN1a data is accepted, it means that our standards of mass, length, and time change as the big bang expands. Only by measuring the gravitational potential, and correcting our standards to a standard metric, can we adjust past and future measurements to a common metric. This study also finds a physical basis for Mach's principle, in that the apparent index of refraction of the space in which we live is defined by all the mass in the universe. Since this appears to fall off with the distance from center, additional and better SN1a data may enable us to locate our place within the big bang. Contrary to the cosmological principle, our big bang may have structure on the largest scale. The calculations and graphs were produced by MathCad™ V6.0.

## Appendix 1. -- Red shift, 19 point data set used in this paper.

From http://journals.uchicago.edu/AJ/journals/issues/v116n3/980111. On the left is the raw data, Z in the first column and magnitude in the second column. $Z_i$ is $M(i,0)$ and magnitude is $M(i,1)$. The Magnitude conversion to distance formula assumes that published magnitudes, $M(i,1)$, are $m_V - M_V$. The absolute magnitude of a SN1a is taken to be $M_V = -19$.

$$r = 32.61 \times (2.5)^{M(i,1)/2} \quad \text{light years}$$

$$M := \begin{bmatrix} .43 & 42.03 \\ .62 & 43.01 \\ .57 & 42.83 \\ .30 & 40.99 \\ .38 & 42.21 \\ .43 & 42.34 \\ .44 & 42.26 \\ .50 & 42.70 \\ 1.2 & 44.60 \\ .48 & 42.49 \\ .30 & 40.74 \\ .23 & 40.33 \\ .16 & 39.08 \\ .24 & 40.68 \\ .17 & 39.95 \\ .83 & 43.67 \\ .011 & 33.78 \\ .089 & 38 \\ .97 & 44.3 \end{bmatrix}$$

## Appendix 2. GR and MMR compared.

GR (general relativity) is based on constant speed of light plus the covariance principle. In GR, rest mass is assumed constant and standards of mass and length and time are everywhere assumed to be the same in every inertial frame, i.e. standards are portable. With this assumption, covariance simply means that the laws of physics must be written in such a form that they survive the transformation from one coordinate system to another. This leads to a connection between distance and time, the Lorentz transformation, and concludes that gravity distorts the very geometry of space from Euclidean to Riemanian. In GR, gravity is a distortion and is not a force. The standards of M, L, and T are immutable.

MMR (mass-metric relativity) is an alternate theory of relativity, which begins by interpreting the 1960 Pound-Rebka Mossbauer experiment on the apparent weight of photons as finding that rest mass $m_0$ increases under gravity and, of course, accepting the SR conclusion that mass also increases with speed (really increases). Dynamical mass, which appears in equations of motions, increases with speed and gravity and is not a constant.

$$m = m_0(\gamma/\alpha) \qquad [A1]$$

where $\gamma$ has the usual SR meaning of $1/\sqrt{1-\beta^2}$ where $\beta=v/c$ and where $1/\alpha$ measures the gravitational potential.

$$1/\alpha \equiv \exp(GM/rc^2) \approx 1+GM/rc^2 \qquad [A2]$$

Any change in rest mass leads to isotropic changes of the metric. By quantum mechanics, any increase of the mass of an electron will cause a physical meter stick to shrink isotropically and an atomic clock will run faster. Covariance is denied, since the meter sticks used to construct coordinate systems shrink in the presence of speed and/or gravity. An "outside" observer may have a different metric than that at the site of an experiment. The outside observer infers that the meter sticks shrink, isotropically, by $\alpha/\gamma$, and that the clocks there have shorter periods by this same factor. Locally, the experimenter reports the speed of light as having the same numerical value, c, using his shorter meter sticks and his faster clock. When the outside observer looks at the same experiment, he finds the meter sticks shrunken but uses his own clock. And so, he finds all speeds under gravity seem to him smaller, by $\alpha/\gamma$. A gravitational red shift causes the inferred faster clock time to slow, by $1/\alpha^2$, when the time is reported to the outside observer.

When one recognizes these changes of metric, with speed and gravity, Newtonian gravity correctly accounts for the classical tests of GR including the Shapiro time delay, the bending of starlight, and the rate of precession of the perihelion of Mercury. The first two of these recognize that the change of metric under a gravitational potential is equivalent to assigning an index of refraction to empty space,

$$n = 1/\alpha^2 \equiv \exp(2GM/rc^2) \qquad [A3]$$

Upon applying Snell's law of refraction, one easily finds correct answers.

It may seem strange that $n = 1/\alpha^2$, instead of $n = 1/\alpha$. The usual definition of index of refraction is the ratio of c to the speed of light c* in a medium, n=c/c*, and this is $1/\alpha$. But this does not include all the "big bang" consequences, which change the metric as was discussed on page 9. There is a change of metric, $1/\alpha$, according to the present radius R. There is an additional change of metric, $1/\alpha$, according to the present mass M. The net result is that $n = 1/\alpha^2$. This "n" is the ratio of time for a light signal to move between two geometric points, with and without gravity, as seen by an observer sited elsewhere. We use light to measure distance, in the SN1a data analysis, and the apparent or "optical" distance increases by $n = 1/\alpha^2$. We have essentially re-defined index of refraction as optical distance divided by geometric distance. Notice that this index of refraction is not c/c*.

The dimensionless gravitational factor,

$$A = (GM/rc^2) \qquad [A4]$$

plays a major role in relativistic physics. The GR Schwarzschild metric is

$$ds^2 \approx dr^2(1+2GM/rc^2) + \text{additional terms in } \omega \text{ and } t \qquad [A5]$$

In earth gravity at sea level, A = $1.4 \times 10^{-9}$ and is not expected to change the metric appreciably. But "A" is much larger if we include all the matter in our big bang. For example, with conventional cosmological values for M=$6 \times 10^{52}$ kg, and R=cT where T=$18 \times 10^9$ years, one calculates from [A4] that A = 0.263 and is not negligible. In this study, we recognize the gravitational potential derived from all the matter in the big bang and seek to measure it using the SN1a data. Assuming a constant spatial density over this spherical volume, the gravitational potential within the big bang is:

$$V = (GM/2R^3)(r^2 - 3R^2) \qquad [A6]$$

This means that we live in a metric, shrunken with respect to the metric "elsewhere" where A=0, by a factor appreciably different from "1.000000". In GR, this GM/R term is not interpreted in terms of gravitational potential. Instead, a single mass "M" is solely responsible for the gravitational acceleration on a test mass and it is this acceleration that GR deals with, using SR, in a series of inertial frames which move with the test mass. Although [A5] contains GM/r, and looks like a gravitational potential, GR does not consider it as such nor does it take into account the gravitational potential due to the big bang which is a billion times greater.

In this paper, the weighted average finds A=0.293. With this, and using the exponential form, n=2.41, M=$6.03 \times 10^{52}$ kg, and R=$16.24 \times 10^9$ light years (in our metric, at this time).

A major difference between GR and MMR is that the metric of GR is inherently anisotropic. In GR, the measured radial distance from a large mass becomes increasingly non-linear as r decreases, but the two orthogonal distances do not. MMR holds that the change of metric is isotropic, by nominally the same factor that GR finds for the radial direction. In the MMR view, and in consequence of this study, we now live and work in a space in which is optically more dense than that "elsewhere" by the factor n = $1/\alpha^2$ = 2.41. And, "n" decreases with time. So, when and where in the world (or beyond) should we define our standards of mass, length, and time? Since we are interested in exploring the cosmos, the big bang in which we live, and since the metric is expected to vary according to the time since the big bang and also how far we are from center, it seems appropriate to define standards of M, L, and T as they would appear "elsewhere" where gravitational influences are not present. In other words, the metric where we live is changing according to the time since the big bang and where we are located. The metric "elsewhere" does not.

Once we recognize this change of the metric, it simplifies the task of dealing with cosmic events occurring long ago and far away.  Referenced to standards "elsewhere", we have a solid basis for the applicable metric both here and also at the place and time where each SN1a erupted and sent its signal towards us.   This allows us to compensate for the changing metric of the space through which the light signal traveled to us.  We also have a basis for extrapolating our metric to future times.

Although we seem to be forever precluded from traveling outside our big bang, if only because its boundaries are receding from us at c, it is still interesting to think on what life might be like "elsewhere" or, eventually, here on earth.   In the long history of the development of life on earth, the metric differed and this could have affected the evolutionary process.

The gpb experiment, now taking data in 2004, will ask hard questions of GR.  Is there really an anisotropic distortion of space, by earth gravity, leading to a Lense-Thirring spin-orbit coupling with the on- board gyroscopes?  Will the gyroscope axes slowly tilt towards their motion in polar orbit about earth?  MMR predicts the absence of  Lense-Thirring precession, and predicts the in-plane tilt will be opposite to the motion.  Time will tell.

In summary, GR holds to the concept of portable and immutable standards of mass, length, and time and relies on a constant c and on the covariance principle.  It works, so far, correcting the Newtonian predictions which had been shown to fail when examined closely.  MMR finds that mass, rest mass, increases with gravity and decreases with time since the big bang.  It accepts that the mass used in orbital calculations is not  constant, but increases with speed and with gravity.  There are metric consequences for this increase of mass, with meter sticks shrinking and clocks ticking faster.  MMR denies covariance, precisely because the metric can differ in different inertial coordinate systems,  whether moving at constant speed with respect to one another, or in different gravitational potentials, or both..  With these changes, Newtonian mechanics works and replicates the classical tests of GR.  GR is able to account for the Shapiro time delay and for the deflection of starlight during a solar eclipse.  Both these tests involve the gravitational effects arising from a single compact mass.  But, GR does not seem to be useful in dealing with the metric effects of gravity where the gravitational potential arises from a uniform distribution of mass as in our big bang. MMR is a scalar theory, and its extension to the distribution of mass in our big bang is easy. Given its success in treating the Shapiro time delay and the deflection of starlight, this application of MMR to the cosmos seems plausible.

Further references on MMR are (arXiv.org/abs/physics/0012059 2000), and R.L. Collins, *RUBBER RULERS AND LACKADAISICAL CLOCKS SIMPLIFY GRAVITY,* Infinity Publishing.com, Haverford, PA, 2003.

## Appendix 3. Age depends on the clock metric.

How much time has elapsed since the big bang, as measured by a clock sited on earth? A lot! As one goes back to the time of the big bang, the metric becomes very large and any real clock runs extremely fast. It is difficult to account properly for the changing metric of time if we begin at the big bang. To avoid infinities, we instead first look at the present time as measured by an "earth clock" with fixed metric. We think the big bang was 16.2 billion years ago, meaning that our clock currently reads 16.2 billion years and counting. We then consider an event in our past, such as the 4.4 billion year age of certain rocks dated by isotopic analysis. Since real clocks, such as the isotopic changes within the rocks, ran faster then, we need to account for this change of metric and expect to find a smaller age (in our present fixed metric).

The following graph shows how we normally think of an "earth clock", reading t. It counts time linearly, in our present (constant) metric ($e^{3A}$), since the big bang. The red line shows this time, less 16.2 billion years so that it reads zero time now. Conventionally, we would think this meant that the ancient rock was created at 11.8 billion years after the big bang.

However, the metric of time for any real clock varies as $\exp(3AT^2/t^2)$ where T is now 16.2 billion years. The real clock ran faster, in the past, as shown by the blue line. TE accounts for the changing metric, which is $-\exp(3AT^2/t^2)+\exp(3A)$, and reads zero at the present time. Upon adding this correction, the blue line shows how a real clock would count time. The isotopes inside the rock constitute a real clock, and the rock was formed when this blue line reached 4.4 billions of years ago. By the earth clock, the rock was formed only 2.9 billions of years ago.

One might reasonably ask, where can one place a real clock so that it will not count time in such a non-linear fashion? We can't get there, physically, but we can imagine placing the clock well outside our big bang where it is away from the changing gravitational potential that makes "earth" clocks run at different rates. According to that clock, unaffected by gravitational potential, time proceeds at a uniform rate. At the present time, where earth clock is reading 16.24 billion years since the big bang, the "elsewhere" clock reads 6.76 billion years.

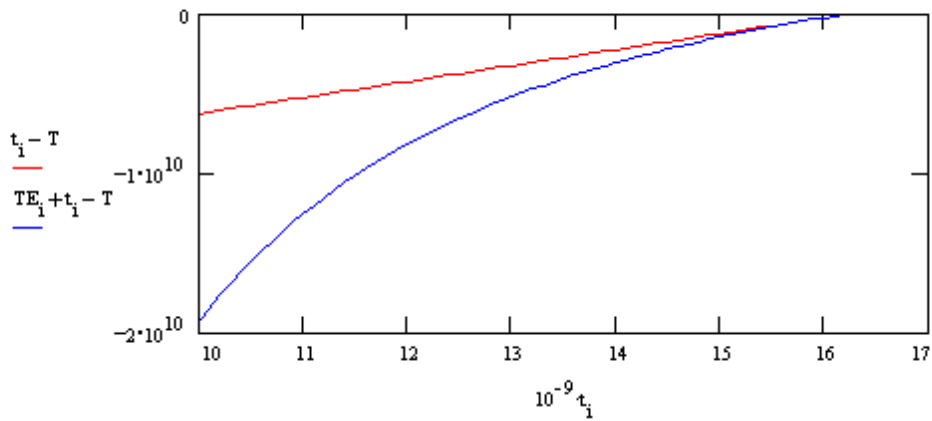

$i := 100..162$

$t_i := i \cdot 10^8$

$T := 16.2 \cdot 10^9$

$A := 0.293$

$TE_i := \int_{t_i}^{T} \left( e^{3 \cdot A \cdot \frac{T^2}{t^2}} - e^{3 \cdot A} \right) dt \cdot (-1)$

$TE_N + t_N - T = -4.439 \cdot 10^9$

$N := 133$

Real (blue) and fixed metric (red) clock readings, synchronized at present.